# Uniform Semiclassical Wavepacket Propagation and Eigenstate Extraction in a Smooth Chaotic System

Daniel Provost and Paul Brumer

*Chemical Physics Theory Group, Department of Chemistry, University of Toronto, Toronto, Canada M5S 1A1*

A uniform semiclassical propagator is used to time evolve a wavepacket in a smooth Hamiltonian system at energies for which the underlying classical motion is chaotic. The propagated wavepacket is Fourier transformed to yield a scarred eigenstate.

Semiclassical mechanics attempts to elucidate and utilize the relationship between classical dynamics and its quantum counterpart in the small $\hbar$ regime [1]. For bound conservative Hamiltonian systems, two goals may be identified: to develop useful methods for the semiclassical propagation of wavepackets and to obtain estimates for the quantum eigenvalues and eigenfunctions using classical trajectories. Both have been achieved for classically integrable systems [2]. For classically chaotic systems, however, the situation is still unresolved.

For the case of eigenfunction or eigenvalue determination, most semiclassical studies of bound Hamiltonian systems have been done in the energy domain. After all, the eigenvalues and eigenfunctions are time independent quantities and should therefore be related to classical manifolds that are time invariant. When the corresponding classical system is integrable, the classical motion is restricted to tori and their quantization produces good semiclassical estimates for the eigenvalues and the eigenfunctions [2]. This is in contrast to the case where the classical system is chaotic: the only time invariant structures other than the energy shell are then the periodic orbits. Consequently, periodic orbits are at the center of the semiclassical efforts in the energy domain for chaotic systems [3]. For example, Gutzwiller's trace formula [4] uses the periodic orbits to explain oscillations in the density of states and a similar expression due to Bogomolny [5] explains the accentuations and attenuations in the average coordinate probability density about the classical periodic orbits, the so-called scars.

At present, however, we cannot obtain individual eigenvalues or eigenfunctions for generic chaotic Hamiltonians using periodic orbit theory. There are two main reasons for this. The first is numerical; to obtain an energy resolution of $\epsilon$ requires all periodic orbits with period $\tau < \tau_{max} \sim 2\pi\hbar/\epsilon$, a quantity which grows exponentially in the chaotic regime [6]. Because of this exponential proliferation of contributing orbits, single eigenvalues can be obtained only for very special systems. The second reason is theoretical and deals with the possible divergence of the semiclassical expressions as we try to resolve an eigenvalue or eigenstate. This divergence can be traced back to the non-commutativity of the limits $\hbar \to 0$ and $\tau_{max} \to \infty$.

Similarly, considerable difficulties abound in attempting semiclassical propagation in chaotic systems over reasonable time periods. For example, Van Vleck–Gutzwiller type propagators [2,7] require trajectories satisfying two-point boundary value conditions. Chaotic dynamics makes finding such trajectories extremely difficult. Hence, generally useful long time propagators are sorely needed to study chaotic dynamics. The availability of such methods would also aid in finding desired eigenvalues and eigenfunctions. In particular, if the propagation time is comparable to the Heisenberg time, $\tau_H = 2\pi\hbar/D$, where $D$ is the mean level spacing, then Fourier transforming the propagated wavepacket exposes the semiclassical eigenvalues and eigenfunctions.

Despite the inherent difficulties, some progress has recently been made toward developing useful propagation methods in the chaotic regime. For example, Tomsovic and Heller propagated wavepackets [8], using the Van Vleck–Gutzwiller propagator and extracted a scarred eigenstate of the stadium billiard [9]. In doing so they propagated past the so called log time [10], $t_{log} \sim \frac{1}{\lambda} \log N_{TF}(E)$ where $\lambda$ is the largest Lyapunov exponent of the chaotic flow [11] and $N_{TF}(E)$ is the number of Planck cells in the phase space volume enclosed by a shell of energy $E$. This time was expected to limit the utility of nonuniform propagators of the Van Vleck–Gutzwiller type [7] since it was thought to be accompanied by the exponential proliferation of caustics. However, Schulman [12] has recently shown that their success in "breaking the log time barrier" arises from pathologies of the particular system chosen, the stadium billiard, where most caustics are found near the walls. For more realistic systems it is unclear whether non-uniform semiclassical expressions can be used for sufficiently long times.

In this letter we use a uniform semiclassical propagator to get the time evolution of a wavepacket in a smooth Hamiltonian system at energies for which the underlying classical motion is chaotic. By uniform we mean that there are no caustics (singularities) in our expressions. We then Fourier transform this propagation to obtain a scarred eigenstate. The method is straightforward and has no caustics–based time limitations. Furthermore, trajectories satisfying two–point boundary values conditions are not required.

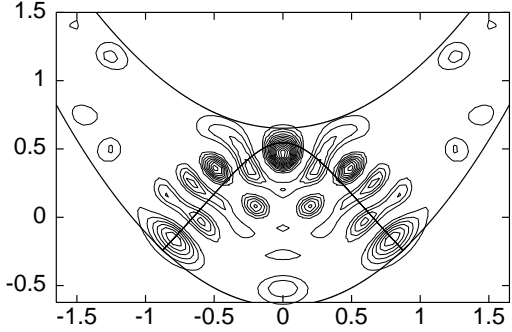

FIG. 1. Contours of the coordinate space probability density of the 42nd even (in $x$) eigenstate, showing striking scarring by a symmetric libration. The equipotential contour at $E = 0.4249$, the energy of this eigenstate, is also shown.

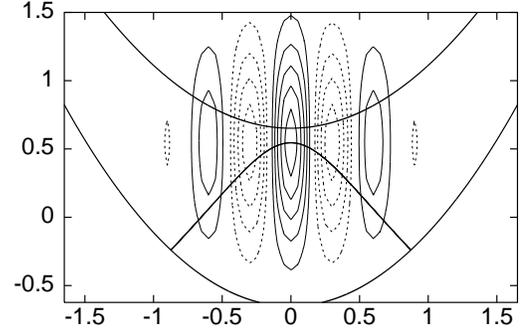

FIG. 2. The coordinate representation of the real part of the coherent state to be propagated. Solid contours are positive, dashed are negative. The contour spacing is 0.2 in Figs. 2, 3, 4, and 5.

To obtain the time evolution of a wavepacket $|\Psi\rangle$ we consider the following expression:

$$\langle\Psi|e^{-i\hat{H}t/\hbar}|\mathbf{q}_i\rangle = \int d\mathbf{p}_f \langle\Psi|\mathbf{p}_f\rangle\langle\mathbf{p}_f|e^{-i\hat{H}t/\hbar}|\mathbf{q}_i\rangle \quad (1)$$

Then $\Psi(\mathbf{q}_i, t) = \langle\Psi|e^{-i\hat{H}(-t)/\hbar}|\mathbf{q}_i\rangle^*$. A semiclassical approximation for Eq. (1) is obtained by replacing $\langle\mathbf{p}_f|e^{-i\hat{H}t/\hbar}|\mathbf{q}_i\rangle$ with a Van Vleck–Gutzwiller type of expression [13]:

$$\langle\mathbf{p}_f|e^{-i\hat{H}t/\hbar}|\mathbf{q}_i\rangle_{sc} = \frac{1}{(2\pi\hbar)^{n/2}} \sum \frac{1}{|\sqrt{\det(D)}|} \times \exp\{i\phi/\hbar - i\nu\pi/2\} \quad (2)$$

The sum is over all classical trajectories that connect $\mathbf{q}_i$ with $\mathbf{p}_f$ in a time $t$; the sum is indexed by the initial momenta $\mathbf{p}_i$ such that

$$\mathbf{q}(0) = \mathbf{q}_i, \quad \mathbf{p}(0) = \mathbf{p}_i, \quad \text{and} \quad \mathbf{p}(t) = \mathbf{p}_f. \quad (3)$$

The matrix $D$ is one of the four stability matrices which describe how trajectories in the immediate vicinity of a given classical trajectory behave:

$$\begin{pmatrix} \delta\mathbf{q}(t) \\ \delta\mathbf{p}(t) \end{pmatrix} = \begin{pmatrix} A & B \\ C & D \end{pmatrix} \begin{pmatrix} \delta\mathbf{q}(0) \\ \delta\mathbf{p}(0) \end{pmatrix} \quad (4)$$

The function $\phi(\mathbf{q}_i, \mathbf{p}_i, t)$ is a generator of the classical motion and is given by

$$\phi(\mathbf{q}_i, \mathbf{p}_i, t) = -\mathbf{p}(t)\mathbf{q}(t) + \int_0^t (\mathbf{p}(\tau)\dot{\mathbf{q}}(\tau) - H)d\tau. \quad (5)$$

Finally the index $\nu[\mathbf{q}_i, \mathbf{p}_i, t]$ is given by [13,14]:

$$\nu[\mathbf{q}_i, \mathbf{p}_i, t] = \frac{1}{\pi} \lim_{\epsilon \to 0^+} \arg\{\det(D + i\epsilon B)\} \quad (6)$$

with $\nu[\mathbf{q}_i, \mathbf{p}_i, t = 0] = 0$. The matrices $D$ and $B$ are defined by Eq. (4). It can be shown that Eq. (6) is equivalent to directly derived Maslov indices [15].

Inserting the semiclassical expression given in Eq. (2) into Eq. (1), we obtain

$$\langle\Psi|e^{-i\hat{H}t/\hbar}|\mathbf{q}_i\rangle_{sc} = \frac{1}{(2\pi\hbar)^{n/2}} \int d\mathbf{p}_i \langle\Psi|\mathbf{p}(t)\rangle |\det(D)|^{1/2} \times \exp\{i\phi/\hbar - i\nu\pi/2\} \quad (7)$$

where $\mathbf{p}(t)$ is the momentum resulting from propagation beginning at $(\mathbf{q}_i, \mathbf{p}_i)$. To obtain Eq. (7) we made the change of integration variables from $\mathbf{p}_f (= \mathbf{p}(t))$ to $\mathbf{p}_i$. Note that this expression has no singularities. In fact, although expression (2) develops caustics (i.e. singularities caused by the focussing of classical trajectories) as time progresses, Eq. (7) is uniform [13]. This expression is in the form of an initial value representation [14–16], i.e. it allows propagation of trajectories starting at $(\mathbf{q}_i, \mathbf{p}_i)$ and does not require [as does, e.g. Eq. (2)] trajectories that satisfy two–point boundary conditions.

We use Eq. (7) to propagate a coherent state (Gaussian wavepacket) in the following smooth Hamiltonian system:

$$H = \frac{1}{2}(p_x^2 + p_y^2) + 0.05x^2 + \left(y - \frac{1}{2}x^2\right)^2. \quad (8)$$

In contrast to billiard systems, this Hamiltonian has a smooth potential and is generic in the sense that the energy cannot be scaled away. Its classical dynamics have been studied extensively [17] and it has been used, in the energy domain, to semiclassically study the scarring of wavefunctions [18]. The exact quantal wavefunctions and energies were obtained as in [18], with $\hbar = 0.05$. In Fig. 1 we show the coordinate probability density of the 42nd even (in $x$) eigenstate; it has energy $E = 0.4249$. We also show the simple symmetric libration that scars it extensively.

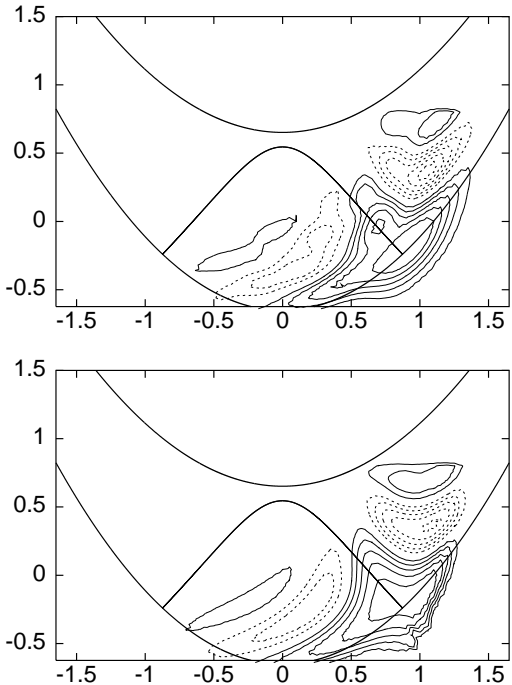

FIG. 3. The real part of the wavepacket propagated for a time $t = \tau/4$, where $\tau$ is the period of the symmetric libration. Top: exact quantal propagation. Bottom: semiclassical propagation.

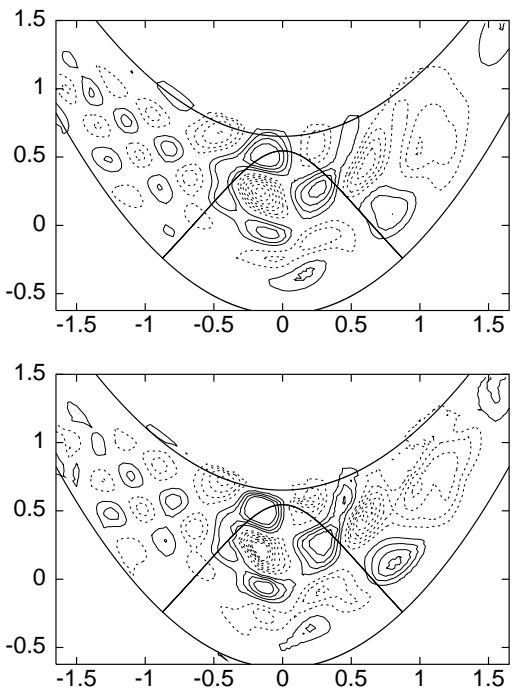

FIG. 4. Same as Fig. 3 but for $t = \tau$.

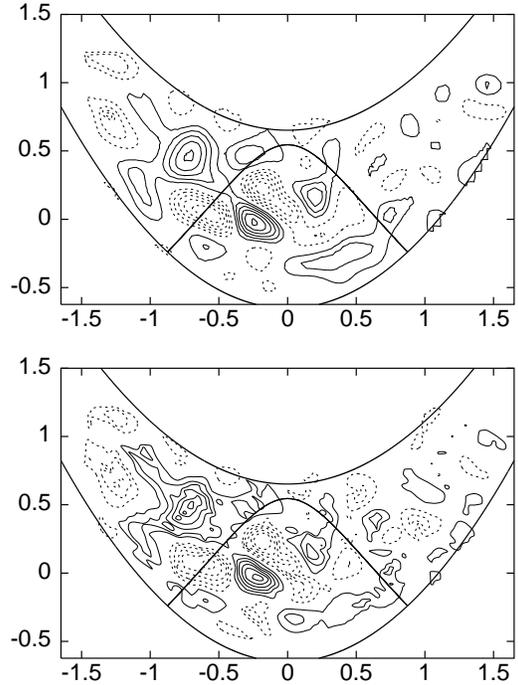

FIG. 5. Same as Fig. 3 but for $t = 2\tau$.

Here we consider propagation of a wavepacket in this classically chaotic regime, extracting this eigenstate semiclassically. To do so we we center a coherent state on this periodic orbit; the coherent state we chose has the following coordinate representation:

$$\langle \mathbf{q}|\Psi\rangle = \left(\frac{1}{\pi b^2}\right)^{1/2} \exp\left\{-\frac{(\mathbf{q}-\bar{\mathbf{q}})^2}{2b^2} + \frac{i\bar{\mathbf{p}}}{\hbar}(\mathbf{q}-\bar{\mathbf{q}}/2)\right\} \tag{9}$$

with $b = 0.5$ and $(\bar{\mathbf{q}}, \bar{\mathbf{p}})$ is the point where the symmetric libration intersects the $y$ axis. In Fig. 2 we show the real part of the coordinate representation of this coherent state, chosen large in coordinate space so that it is well localized in momentum space. This feature allows us to search out the relevant regions in momentum space where Eq. (7) must be evaluated and to carry out the integration with low order quadrature methods. Note that we do not evaluate Eq. (7) by stationary phase and hence avoid the root search problem and caustics associated with coalescing trajectories.

Our calculations of the semiclassical wavepacket propagation is illustrated in Figs. 3 – 5 and are compared with the exact propagation. The agreement between the exact quantal and semiclassical pictures is excellent, even past the log time which we estimate to be $t_{log} \sim 1.5\tau$, where $\tau$ is the period of the symmetric libration. In Fig. 6 we show the Fourier transform of our exact and semiclassical wavepacket propagations, with $\tau_{max} = 2.5\tau$. Since we also have the wavepacket propagations backwards in time, we expect to get an energy resolution equivalent to a propagation time of $5.0\tau$. This is smaller than the Heisenberg time, which is $\tau_H \sim 8.0\tau$ at $E = 0.4249$. However due to special wavepacket we chose, we are able to resolve most of the eigenstate and agreement between the semiclassical and quantum results is excellent. It is important to realize that a similar calculation in the energy domain using Bogomolny's formula [18] would not produce the nodal structure which is seen to exist along the libration, even if all the periodic orbits with periods up to $\tau_H$ were included.

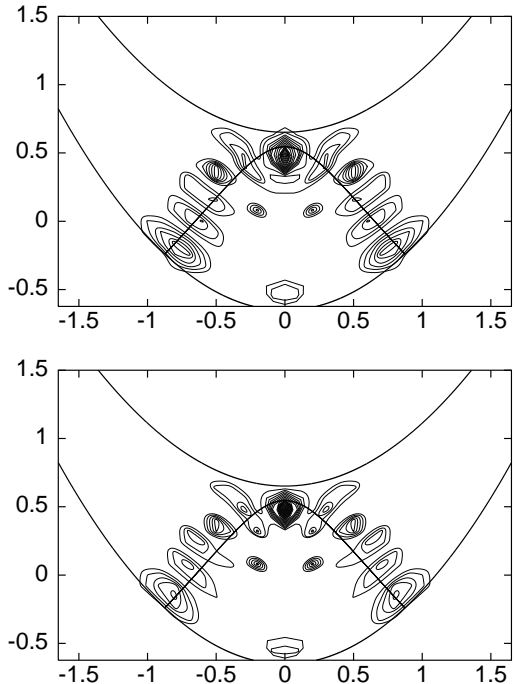

FIG. 6. The Fourier transform of the wavepacket propagation with $E = 0.4249$ and $t_{max} = 2.5\tau$. Top: from exact quantal propagation. Bottom: from semiclassical propagation.

The resultant figures clearly demonstrate that this uniform propagation method is capable of accurate propagation in chaotic systems over sufficiently long times to extract accurate eigenstates. Further studies over longer time periods and for larger systems are well warranted and are in progress.

## ACKNOWLEDGMENTS

This work was supported in part by funds provided by the Ontario Laser and Lightwave Research Center and the U. S. Office of Naval Research under contract number N00014-90-J-1014.